# Light-activated memristor by Au-nanoparticle embedded HfO$_2$-bilayer/p-Si MOS device

Ankita Sengupta[1], Basudev Nag Chowdhury[1], Bodhishatwa Roy[1], Biswarup Satpati[3], Satyaban Bhunia[3], Sanatan Chattopadhyay[1,2,*]

[1]Department of Electronic Science, University of Calcutta, Kolkata, India

[2]Centre for Research in Nanoscience and Nanotechnology, University of Calcutta, Kolkata, India

[3]Saha Institute of Nuclear Physics, A CI of Homi Bhabha National Institute, Kolkata, India

*Email: scelc@caluniv.ac.in

**Abstract**

The current work proposes a novel scheme for developing a light-activated non-filamentary memristor device by fabricating an Au-nanoparticle embedded HfO$_2$-bilayer/p-Si MOS structure. Under illumination, the electrons in such embedded Au-nanoparticles are excited from d-level to quantized s-p level and are swept out on application of an appropriate gate bias, leaving behind the holes without recombination. Such photogenerated holes are confined within the nanoparticles and thus screen the external field to lead to a memristive effect in the device. The phenomenon is experimentally observed in the fabricated Pt/HfO$_2$-(layer-II)/Au-NPs/HfO$_2$-(layer-I)/p-Si devices, where such memristive effect is activated/deactivated by light pulses. The memory window and high-to-low resistance ratio of the device are obtained to be ~1 V and ~10, respectively, which suggest the performance of a standard state-of-the-art memristor. Further, the present device offers a voltage-sweep-endurance up to at least 150 cycles and the memory retention up to ~10$^4$ s. Such a device concept can be extended for a combination of different nanoparticles with various dimensions and dielectric layers to optimize their memristive effect for achieving CMOS-compatible memory devices with superior reliability.

**Keyword:** Resistive switching; Memristor; Light-activated memristors; Au-nanoparticle; Si MOS.

Resistive switching memory devices, demonstrating their potential application as non-volatile memory [1] along with scalability and low-power advantages [2], have recently exhibited further interest as a promising fundamental unit of beyond-von Neumann computing systems with 'in-memory' architecture [3]. Alongside being memory units, such devices have also shown extensive applications in data encryption [4], RF operations for mobile communications [5, 6], cryptographic systems including true random number generators [7], 5G network [8] and also as terahertz switches [8, 9]. Several classes of materials including metal-oxides, ferroelectrics, phase change materials (PCMs), magnetic and 2-D materials, as well as nanoparticle (NP) embedded insulators are explored to realize resistive switching behavior [10-22]. Such materials, being exploited as active memory component in metal-insulator-metal device structures, have been claimed to show excellent endurance and retention property along with significantly fast reading/writing speed [23, 24].

However, identifying the key challenges associated with variability and reliability issues, a recent work observes that majority of such claims are highly inaccurate and unreliable due to the lack of adequate data support [25]. It is imperative to note that the concern about such cycle-to-cycle and device-to-device variability and output reliability is essentially originated from the physical mechanism of memristive effect. In fact, the resistance of memristive materials, on application of certain bias, suffers a transitory change which is retained up to a much smaller voltage, and thereby indicates that such property can be utilized for memory applications. Such memorizing capacity is attributed to the rearrangements of atoms, metallic ions or vacancies, induced by electrical, thermal or optical pulses, which leads to the formation/annihilation of a conductive filament [26-29]. It is apparent that such induced rearrangements are statistical in nature thereby making the conductive path formation inherently stochastic, resulting in the output variability. Furthermore, amongst such memristive materials, the PCMs suffer from stability issues regarding their amorphous and crystalline state transitions [30], the magnetic materials exhibit small high-to-low resistance ratio [1], and the device-to-device variability is very high for ferroelectrics [1, 31]. Also, most of the proposed materials and devices exhibiting memristive effects are not compatible with the state-of-the-art complementary metal-oxide-semiconductor (CMOS) technology.

In this context, the present work proposes a novel scheme to achieve a light-activated memristive effect without the formation of any conductive filament, and thereby reduces the possibilities of cycle-to-cycle variability or reliability. For such purpose, a Pt/HfO$_2$-(layer-II)/Au-NPs/HfO$_2$-(layer-I)/p-Si MOS device is fabricated in the present work as a proof-of-the-concept of non-filamentary resistive switching memory, which is activated/deactivated by incident light. Such device structure is schematically represented in Fig. 1(a), while the physical mechanism for memristive operation is illustrated in Fig. 1(b) by showing the band alignments in a memory cycle, the top-left one depicting the equilibrium band diagram at dark condition. If such device is illuminated with light of relevant wavelength, electrons in the Au-NPs can be excited from its occupied d-band to the unoccupied s-p band above Fermi level near X and L-points, leaving behind an equivalent hole in the d-band [32-34]. Nevertheless, such photogenerated electron-hole pairs readily recombine through radiative or non-radiative processes at equilibrium condition (top-right of Fig. 1(b)). However, on application of a positive bias at the metal terminal, the MOS would exhibit a band bending consecutively toward 'depletion' and 'inversion' modes. During the increase of such bias, when the potential barrier between s-p band of Au-NPs and conduction band of HfO$_2$-(layer-II) reduces so as to allow the photogenerated electrons to be transported from Au-NPs to metal (bottom-right of Fig. 1(b)), the device would observe a significant rise in the photocurrent. At such condition the device will transit from its high-resistance state (HRS) to the low-resistance state (LRS). If the voltage is increased further for lowering such Au-NP/HfO$_2$-(layer-II) barrier up to the s-p band energy of Au-NP, or more, the photocurrent would eventually get saturated. It is interesting to note at this point that, the equivalent photogenerated holes at the d-band of AuNPs would be confined due to the barrier of HfO$_2$-(layer-I) valence band, and will behave as positive fixed-oxide-charges in the MOS. Such positive fixed-oxide-charges will induce an electric field in opposite direction to that due to applied voltage thereby screening it (bottom-left of Fig. 1(b)). In such condition, if the applied bias is decreased, such field screening by d-band holes would prevent to recreate the pre-set barrier between HfO$_2$ conduction band and Au-NP s-p band, resulting in the continuation of saturation current through the device. At a much lower voltage such barrier will be redeveloped leading to the recombination of photogenerated electron-hole pairs in Au-NPs, which would then drop down the device current leading to a transition from LRS to HRS. In such condition, the MOS will finally get back to its initial equilibrium condition (top-left of Fig. 1(b)), even under illumination. Therefore, such scheme provides the opportunity to achieve a MOS based non-filamentary resistive switching memory device activated under light illumination.

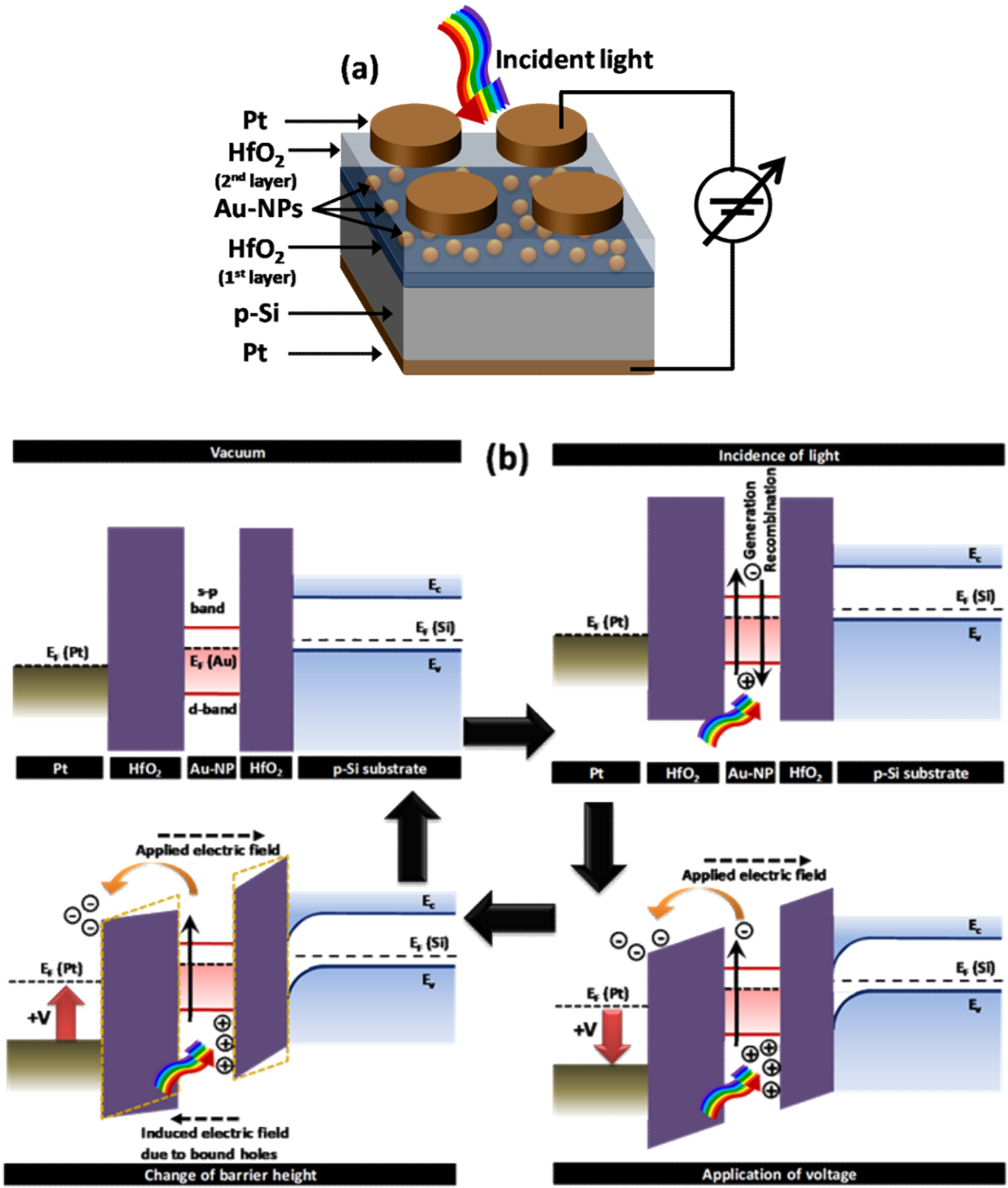

Fig. 1 (a) Device schematic showing Au-NPs embedded in $HfO_2$-stacks on p-Si substrate for light-activated MOS based memory; (b) schematic representation of the device operation illustrating the band alignment for non-filamentary resistive switching, activated/deactivated upon light illumination in one memory cycle: the top-left shows equilibrium under dark condition; top-right depicts photogeneration/recombination within the Au-NPs upon illumination; bottom-right describes the band-bending on applying a positive bias at metal terminal, which in-turn decreases the barrier height between Au-NP and $HfO_2$-(layer-II) leading to transport of photogenerated electrons to metal; and bottom-left illustrates the field screening by confined photogenerated holes inducing an electric field in opposite direction to the applied bias.

A proof-of-the-concept of such scheme is experimentally realized in the present work by fabricating a Pt/$HfO_2$-(layer-II)/Au-NPs/$HfO_2$-(layer-I)/p-Si MOS device, where $HfO_2$ films and Pt contact pads are deposited by RF and DC magnetron sputtering technique, respectively. Prior to the deposition of first $HfO_2$ layer, the p-Si substrate is cleaned following standard RCA-I and RCA-II protocols, followed by dipping it into 1% HF solution for 5 min to remove any residual native oxide layer [35, 36]. Then the sample is placed in sputtering chamber, where the first $HfO_2$ layer is deposited with 110 W for 15 minutes maintaining the base-pressure to be $\sim 5\times 10^{-6}$ mbar. The grown $HfO_2$ film on p-Si substrate is then spin coated with a solution of Au-NPs (for detailed see [SM: S1]) at a speed of 700 rpm. Subsequently, the second $HfO_2$ layer is deposited by sputtering under identical conditions. Finally, Pt is sputtered as top and back contacts of the device, where the top contacts are formed using a patterned shadow mask to provide the area of each device to be $\sim 1.26\times 10^{-3}$ $cm^2$ (*i.e.*, diameter $\sim 400 \mu m$).

Fig. 2(a) shows the FESEM (using Zeiss Auriga) image of surface morphology of the second $HfO_2$ layer (*i.e.*, $HfO_2$-(layer-II)) covering the Au-NPs dispersed on first $HfO_2$ layer (*i.e.*, $HfO_2$-(layer-I)) deposited on p-Si substrate. The image of such Au-NP dispersed first $HfO_2$ layer is shown in the inset of Fig. 2(a). The average diameter of Au-NPs and their average spatial density on the $HfO_2$ film are obtained to be $\sim 13$ nm and $\sim 10^{11}/cm^2$, respectively (for detailed see [SM: S1]). The cross-sectional view of such fabricated device is depicted in the TEM image (using FEI Tecnai $G^2$ F30-ST) of Fig. 2(b) showing each material layer of the device. The thicknesses of first and second $HfO_2$ layers are measured from such TEM image as well as from spectroscopic ellipsometry (for detailed see [SM: S2]), and found to be $\sim 15$ nm each. It can also be observed from Fig. 2(b) that interfacial layers of $\sim 2$ nm thickness are formed between two $HfO_2$ layers as well as at the $HfO_2$/Si junction, which are possibly $HfO_x$ (x<2) that further contribute to the corresponding oxygen vacancy levels. The TEM analysis suggests dominant electron diffractions from ($\bar{1}$11)-plane of monoclinic $HfO_2$ and (111)-plane of Au-NPs (for detailed see the SAED pattern analysis in **[SM: S3]**). The intensity variation of X-rays emitted from such $HfO_2$/Au-NP/$HfO_2$ interface in the energy dispersive spectroscopy (using EDAX Instrument) measured *in-situ* under TEM set up is plotted in Fig. 2(c), where the elemental existence of Hf, O and Au are confirmed from the respective characteristic emission lines (indexed in Fig. 2(c)). The XRD-profile (using PANalytical X'pert Powder) of the first layer $HfO_2$ sputtered on p-Si substrate, measured using Cu-k$\alpha_1$ source ($\lambda$= 1.540598 Å), is shown in Fig. 2(d), where the diffraction peak at 2$\theta$ = 28.3° (see the inset of Fig. 2(d)) corresponds to the ($\bar{1}$11)-plane of monoclinic $HfO_2$ **[JCPDS No: 78-0050]**. As indicated in Fig. 2(d), all the other peaks obtained in XRD-profile are due to contribution from the Si substrate.

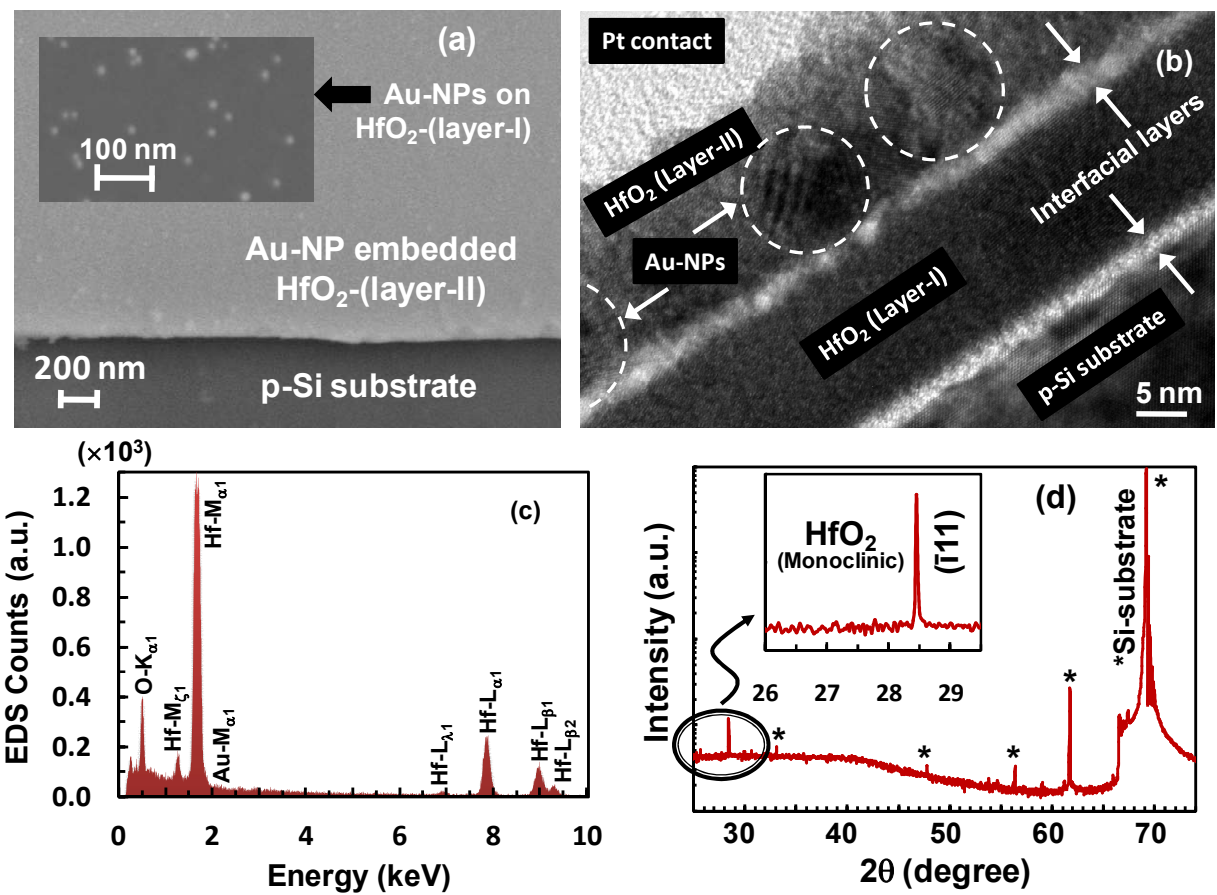

Fig. 2(a) FESEM image of the top-view of $HfO_2$-(layer-II) deposited on Au-NP dispersed $HfO_2$-(layer-I) on p-Si substrate, the inset depicting Au-NPs on $HfO_2$-(layer-I); (b) TEM image of the cross-section of Pt/$HfO_2$-(layer-II)/Au-NPs/$HfO_2$-(layer-I)/p-Si; (c) the plots of energy dispersive

X-ray spectroscopy (EDS) counts with e-beam energy showing relevant characteristic X-ray peaks originated from Hf, O and Au; and (d) XRD profile of the HfO$_2$/p-Si sample along with the inset indicating the relevant crystal plane of monoclinic HfO$_2$.

The photoluminescence (PL) spectrum of Au-NPs embedded within HfO$_2$ layers of the present device is plotted in Fig. 3(a), where multiple peaks are deconvoluted to obtain the associated photo-excitation/de-excitation levels. The PL peaks obtained at ~3.2 eV and ~3.4 eV originate from the transitions related to oxygen vacancy levels $V^0$ and $V^+$ of monoclinic HfO$_2$, respectively [37-40], and are also observed in the two-layer HfO$_2$ film without Au-NPs (for detailed see [SM: S4]). Such vacancies are formed due to the absence of a neutral oxygen atom or an oxygen ion from their lattice positions leaving two electrons or a single electron, respectively, in the vicinity of the corresponding vacant site [37]. Therefore, in the present device such vacancies are attributed predominantly to the interfacial HfO$_x$ layers. The PL peaks at ~2.3 eV, 2.5 eV, 2.7 eV, 2.8 eV and 3 eV, however, are not observed to appear in the two-layer HfO$_2$ film without Au-NPs (see [SM: S4]), and attributed to the radiative transitions in Au-NPs. Such values are also in complete agreement with the previously published reports on photoluminescence peaks of Au-NPs and thin films [41].

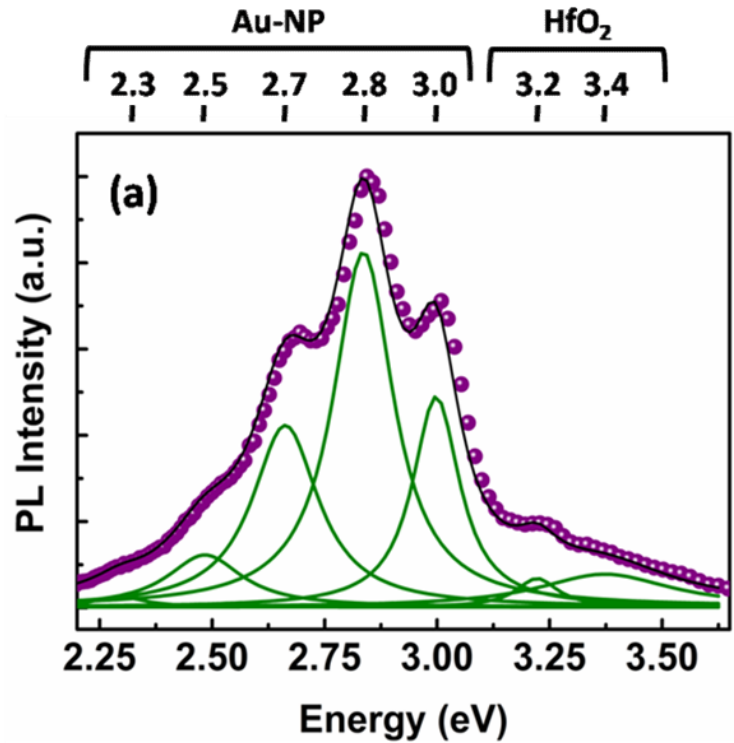

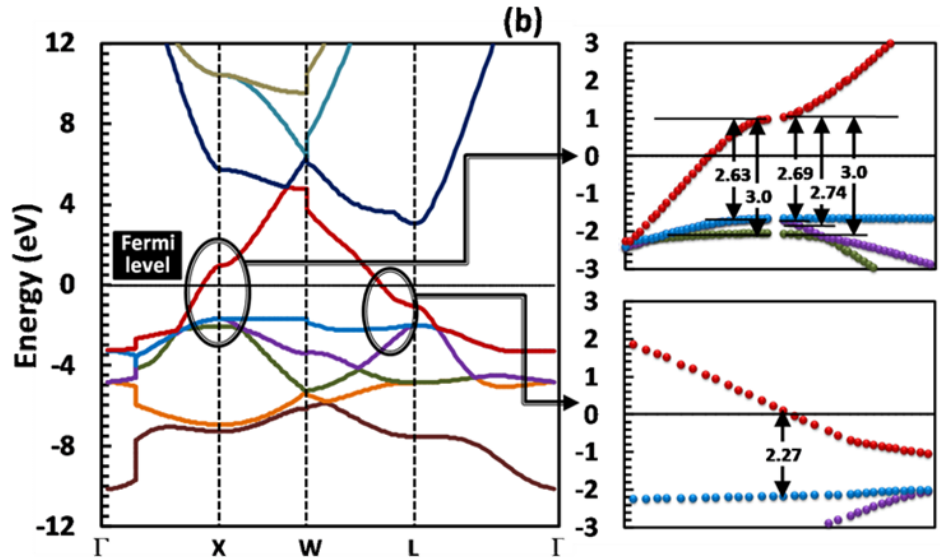

Fig. 3(a) Photoluminescence (PL) intensity with energy emitted from Au-NP embedded $HfO_2$ on Si substrate along with deconvoluted peaks appearing at ~2.3 eV, 2.5 eV, 2.7 eV, 2.8 eV, 3 eV, 3.2 eV and 3.4 eV (symbols: experimental data; green solid lines: simulated deconvoluted peaks); (b) band structure of Au-NP (radius =13 nm) obtained from DFT base first principle calculations, considering electron confinement resulting in momentum discretization leading the possible emission lines at 2.63 eV, 2.69 eV, 2.74 eV and 3.0 eV near X-point (top-right) and 2.27 eV near L-point (bottom-right).

As discussed earlier, such visible photoluminescence in Au thin films and nanoparticles originates from the 6s-p→5d inter-band transitions near X- and L-points **[32-34, 41]**. The corresponding mechanism is described by three possible sequential steps: first, photo-excitation of an electron from the occupied d-band to an unoccupied s-p band above Fermi level leaving a hole in the d-band; second, non-radiative relaxation of the electron to Fermi level; and finally, radiative recombination, predominantly near the van Hove singularity **[32]**. The net photo-response of Au-NPs in such process is claimed to depend mostly on their localized surface states and the degree of quantum confinement **[41, 42]**, which are theoretically investigated by simplified semi-classical approaches **[41, 43]** as well as by attempting to model its bandstructure in the frameworks of time-dependent density functional theory **[41, 43, 44]**. However, it has been predicted that quantum confinement induced energy-momentum discretization in metal nanoparticle bandstructures can significantly improve the photoemission modeling, which is yet to be done [41]. In this context it is worthy to mention that, a recent theoretical model has described band splitting by momentum-quantization in nanotubes validated by experimental observation of emission lines in luminescence study [45]. The model considers spatial confinement of electrons due to reflection from the surfaces of a nanostructure thereby forming standing waves inside it, and leading to discrete values of momentum, which are used in bandstructure calculation by linear combination of atomic orbitals (LCAO) approach [45]. The same model is used in the present work to obtain the discretized momentum in Au-NPs and utilized in density functional theory (DFT)-based first principle calculations to obtain its bandstructure (for detailed see [SM: S5]). The DFT simulations (using Quantum Espresso [46]) are performed using local density approximation (LDA) method, with utrasoft (USPP) exchange-correlation functional for structure relaxation, where the discrete momentum/wave vector inputs are obtained from spherical Bessel zeros as the solution of Schrödinger equation (see [SM: S5]), assuming the shape of nanoparticles to be spherical. The obtained bandstructure of Au-NP (for diameter =13 nm) is plotted in Fig. 3(b), with enlarged plots near X- and L-points showing the possible emission lines. The momentum quantization resulting in disallowed high symmetry points, as can be observed from such plots of bandstructure, gives rise to the possible emission energies to be 2.27 eV near L-point and 2.63 eV, 2.69 eV, 2.74 eV and 3.00 eV near X-point, which are in good agreement with the experimental results. It is also interesting to note from Fig. 3(b) (top-right) that the discrete emission lines near X-points are more close to van Hove singularity (compared to L-point), and thus results in larger PL intensity as can be seen from Fig. 3(a). Further, the slight differences of theoretical and experimental results may be due to not incorporating the room temperature corrections [45]. It is worthy to mention that without considering such momentum quantization, the bandstructure shows only two emission energies of ~1.8 eV and ~2.4 eV at X- and L-point, respectively (for detailed see [SM: S5]), which are consistent with the previous calculations [32-34].

Electrical characterizations including the measurement of current (using Keithley 2611B) and capacitance (using Keysight E4980AL) of the fabricated Pt/HfO$_2$/Au-NP/HfO$_2$/p-Si devices are performed at dark condition as well as under illumination using a solar-simulator (AM 1.5 G with total light intensity of 1000 W/m$^2$), and the results are plotted in Fig. 4(a). It is apparent from the plots of Fig. 4(a) that in dark condition under a positive bias, the room temperature leakage current of the present MOS device is considerably small (~0.01 A/cm$^2$). However, when light is incident on such device, the current increases for voltages ≥1.7 V and starts saturating (saturation value ~0.33 A/cm²) at ~2.5 V, which indicates the device state transition from HRS to LRS. While applying the voltage reverse sweep, the device, showing saturation current, remains in its LRS until the bias decreases to ~2.0 V, and with further reduction of voltage to ≤0.7 V the device comes back to its original HRS. In the negative voltage region, neither the impact of light nor such memory effect is observed due to majority carrier flow. Further, it is imperative to note that the smooth HRS↔LRS transitions rather than discrete jumps suggest no formation of conductive filaments. Therefore, the proposed Au-NP embedded $HfO_2$ based MOS device exhibits unipolar non-filamentary resistive switching memory characteristics under illumination, providing a memory window of ~1 V and an OFF/ON resistance ratio ($R_{HRS}/R_{LRS}$) of ~10, which are at par to the requirements of state-of-the-art integrated circuits [1].

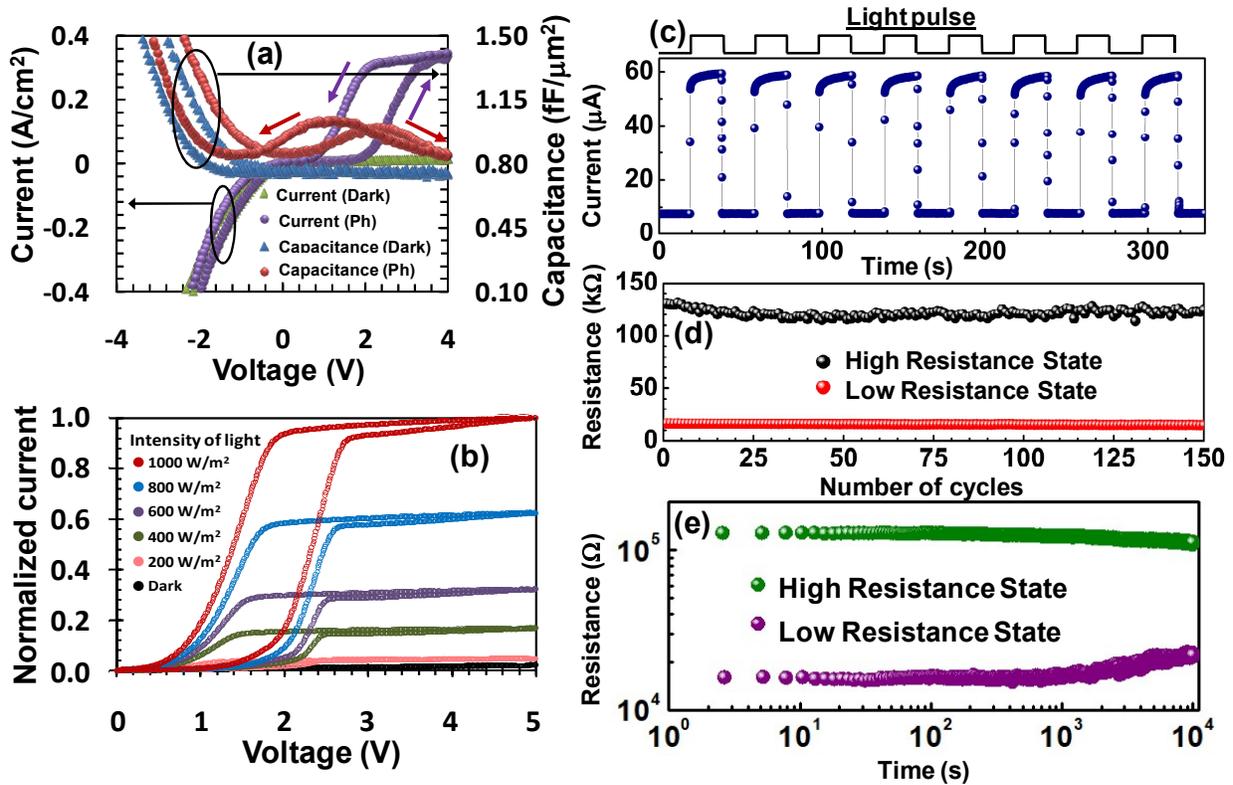

Fig. 4(a) Plots of the current-voltage and capacitance-voltage (at 800 kHz) characteristics of the Pt/$HfO_2$/Au-NPs/$HfO_2$/p-Si device under dark and illumination condition (using solar simulator), showing a unipolar resistive switching behavior upon illumination; (b) the variation in device current with the intensity of incident light; (c) light-activated switching behavior of the device at

1 V, light pulses are applied with a time period of 20 seconds; and the plots of HRS and LRS at 1 V (*i.e.*, effectively at ~0.3 V since the light-activated memory effect appears/remains only at voltages ≥0.7 V): (d) with voltage sweeping cycles for endurance, and (e) as a function of time for retention.

At this point it is worthy to note from the bandstructure of Au-NP in Fig. 3(b) (top-right), that, the momentum-quantized points of s-p band contributing maximum to the photo-generation/recombination at van Hove singularity is ~1.027 eV above the Fermi level, that leads to the $HfO_2$/Au-NP(s-p) barrier height (with an affinity difference of 2.75 eV) for electrons to be ~1.723 eV, which defines the switching voltage for SET, *i.e.*, HRS→LRS, of the device. As discussed earlier, the corresponding photogenerated holes at the d-band of AuNPs screen out the applied field, and thereby retain the band-bending during reverse voltage sweep leading to the memory effect. Hence, the LRS→HRS transition voltage is defined by such photogenerated hole concentration, and thus would predominantly depend on the number density of Au-NPs.

Further, it is worthy to mention that such MOS device without Au-NPs also show a similar rise in photocurrent with voltage due to transitions from oxygen vacancies of $HfO_2$ (along with the contributions from the interfacial layers), although of much smaller value (saturation current ~0.06 $A/cm^2$), and moreover, no memory effect is observed in such devices (see **[SM: S6]**). It also confirms the conceptual proposition of the present scheme of utilizing photogeneration/recombination in Au-NPs for achieving the memory effect. The scheme can be further corroborated with MOS-capacitance study of the device under dark and illuminated condition, where the latter exhibits a characteristic photogeneration peak in both forward and reverse sweep. During forward sweep, such capacitance peak starts diminishing at ~2.5 V where current starts saturating (*i.e.*, SET), while for reverse sweep it rises at ~2.0 V where the current starts reducing (*i.e.*, RESET). Such peaks appearing at accumulation-to-depletion transition region are attributed to the confined holes **[47, 48]**. It is also imperative to mention that the capacitance shown in Fig. 4(a) is measured at a high frequency of 800 kHz confirming that such peak is not originated from the interface traps, which generally cannot respond at such higher frequencies **[47-49]**.

The light-activated resistive switching property of the present device is further investigated by varying the illumination intensity from 0 W/m$^2$ to 1000 W/m$^2$ and plotted in Fig. 4(b). It is apparent from the figure that the saturation current increases with light intensity which also validates the resistive memory observed in the present device to be indeed photo-induced. The exponent ($\eta$) of light intensity in the variation of such photocurrent is calculated from the slope of their logarithmic plot [50-52], and found to be less than 1 for a voltage <1.65 V (*e.g.* $\eta = 0.63$ at 1 V, for detailed see [SM: S7]), which is attributed to significant recombination [50-52] of the photogenerated carriers in Au-NP at such voltages. However, the value of such exponent is obtained to be $\eta = 1$ at ~1.65 V, indicating minimal electron-hole recombination. It is interesting to note that such voltage is close to the empirically observed SET-voltage (~1.7 V), which is in complete agreement with the proposed mechanism in the present work. Further, above such voltage, the value of the exponent increases up to ~2 (at saturation current), which may be attributed to the multiple carrier generations from several possible quantized bands as shown in Fig. 3(b).

The light-activated switching property of the present device is illustrated in Fig. 4(c) showing the transition of device current at 1 V (after SET at 2.5 V) with light pulses (ON/OFF) of 20 s time period, which suggests an excellent repeatability of the optical switching behavior of such device. Although the device shows extremely sharp ON↔OFF transitions, a close inspection reveals that upon switching on the light pulse, after a sharp rise to ~90% of the maximum photocurrent, it exponentially saturates in time slowly (*e.g.*, time constant ~3 s). Such region (*i.e.*, ~90-100% of maximum) may be attributed to the mutually countering effects of recombination and multiple-photogeneration in the Au-NPs. Further, the endurance of such light-activated memory device for 150 cycles is demonstrated in Fig. 4(d) and the memory retention is shown in Fig. 4(e) up to $10^4$ s. The corresponding HRS and LRS are measured at 1 V, *i.e.*, effectively at ~0.3 V, since the light-activated memory effect remains for ≥0.7 V only. It is worthy to mention that the endurance is measured by voltage sweeps (*i.e.*, not by pulse voltage method), which exhibits highly steady HRS/LRS under illumination at least up to 150 full cycles. It is also apparent from the plots of Fig. 4(e) that the memory retention is quite satisfactory for the proposed device. However, such retention property observes a small degradation after continuous measurement for around $10^4$ s, which may be attributed to the resistive heating at the contacts.

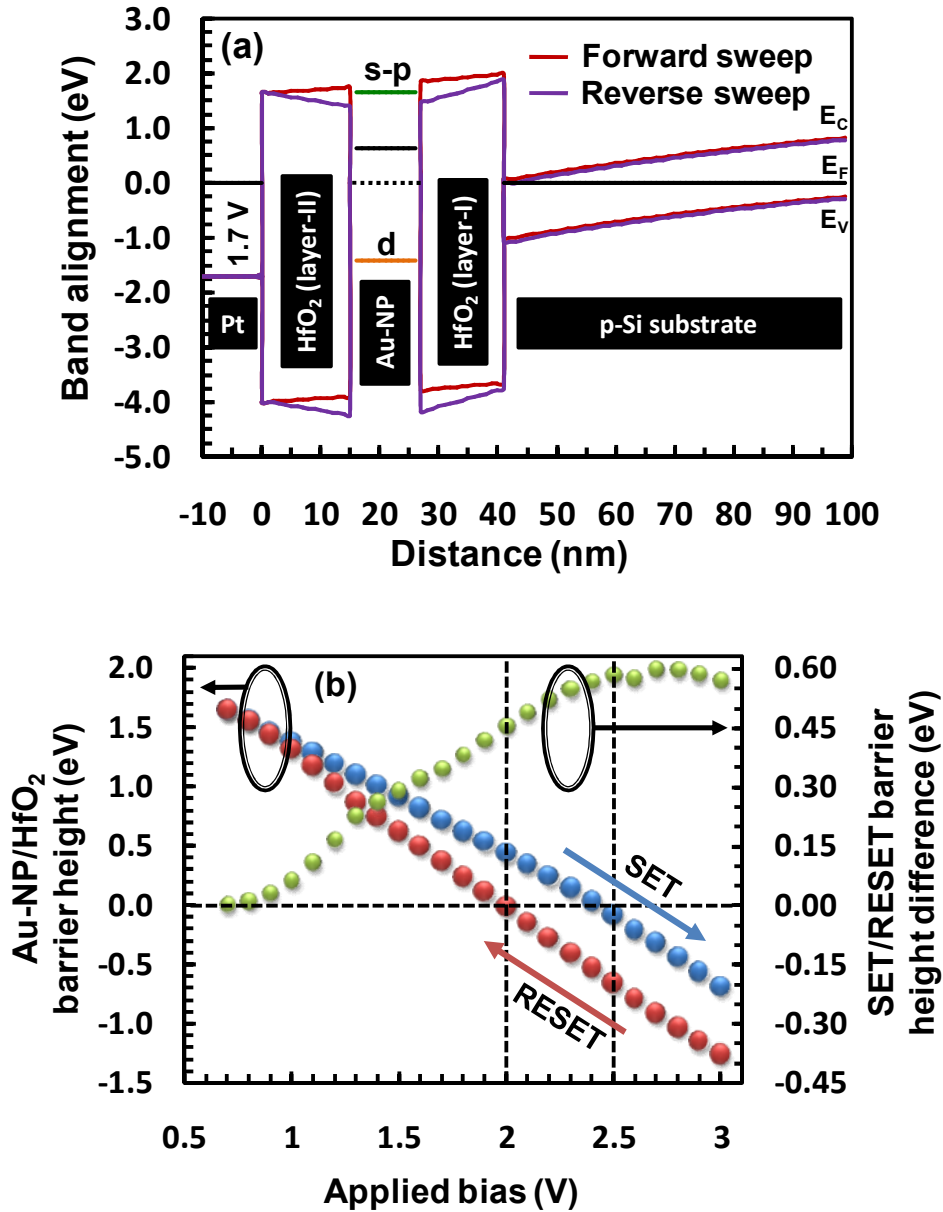

Fig. 5 (a) Simulated band alignment of Pt/HfO$_2$/Au-NP/HfO$_2$/p-Si MOS device during forward and reverse voltage sweeps at 1.7 V (*i.e.*, ~SET-voltage), considering the photogenerated confined holes of Au-NPs to be positive fixed-oxide-charges, their density being calculated from the photocapacitance (of Fig. 4(a)); (b) the plots of barrier height for electrons at Au-NP s-p quantized level and $HfO_2$-(layer-II) with the applied bias, during SET and RESET processes, along with the corresponding SET/RESET barrier height differences.

Further, in order to develop a generalized comprehensive understanding on the proposed mechanism for light-activated non-filamentary resistive memory, originated from the electric field screening by confined holes of Au-NPs inside the $HfO_2$ layers, the effective (Au-NP/ $HfO_2$) barrier height for forward and reverse sweep are calculated from the relevant band alignments. Such band diagrams of the present device structure are simulated using Atlas tool in SILVACO TCAD software (for detail, see **[SM: S8]**). The confined photogenerated hole density at each voltage during forward and reverse sweep is calculated from the experimentally measured capacitance peak under illumination (see Fig. 4(a)), and such confined holes are assumed to behave as corresponding positive fixed-oxide-charges inside $HfO_2$ layers for simulation purpose. Fig. 5(a) represent the simulated band alignment at 1.7 V (*i.e.*, ~SET-voltage) for forward and reverse sweep conditions, where momentum quantized s-p and d-levels of Au-NP are considered according to their maximum contribution to photoluminescence (see the discussion of Fig. 3(a)-(b)). It is apparent from the figure that such s-p quantized level is just below the $HfO_2$-(layer-II) barrier during forward sweep indicating such bias (~1.7 V) to be the SET voltage. However, for reverse sweep condition, such barrier is below the s-p level suggesting a residual device current providing the memory effect. It is interesting to note that the difference in $HfO_2$ band slopes at forward and reverse sweep (Fig. 5(a)) are due to the confined holes in Au-NPs at 1.7 V being ~$2.5 \times 10^9$ /cm$^2$ and ~$4.7 \times 10^{10}$ /cm$^2$, respectively, clearly showing the corresponding field screening difference by such holes. Thus, during forward voltage sweep, the photogenerated electrons of Au-NPs start considerably transporting at ≥SET-voltage due to $HfO_2$/Au-NP barrier lowering and consequently more of the 'un-recombined' holes get confined continuously with increasing the applied bias. However, at the time of reverse sweep such confined holes screen the applied field thereby keeping the barrier lowered and maintaining the electron current up to a much lower value of RESET-voltage, which gives rise to the memristive effect. Such $HfO_2$/Au-NP barrier heights during SET and RESET along with their difference are plotted in Fig. 5(b) (for detail, see **[SM: S8]**). It is evident from the plots of Fig. 5(b) that SET/RESET barrier height difference increases with applied bias due to the 'un-recombined' holes of Au-NPs with a saturation value after ~2.5 V, whereas such difference gets reduced to zero at ~0.7 V below which the device does not exhibit any memory effect.

In conclusion, a novel scheme for light-activated non-filamentary memristor device is proposed and experimentally validated by fabricating a prototype of an Au-NP embedded $HfO_2$/p-Si MOS structure. The fabricated device is observed to provide a memory window of ~1 V and $R_{HRS}/R_{LRS}$ ~10, that satisfies the acceptability standards of the state-of-the-art memristors. The optical switching of such device along with the memory endurance and retention exhibits desired performance standards. The memorizing capacity of such proposed device structure is originated from the 'un-recombined' holes of Au-NPs, confined by $HfO_2$-(layer-I), resulting to an electric field screening in the $HfO_2$-(layer-II). The SET-voltage of such memory device is defined by the Au-NP/$HfO_2$-(layer-II) barrier height whereas the RESET-voltage predominantly depends on the nanoparticle density as fixed charge holders. The other engineering aspects for performance improvement of such device may be the choice of oxides (*e.g.*, $SiO_2$, $ZrO_2$, $Al_2O_3$, $La_2O_3$) for the two layers including combinations of such oxides, different sets of oxide thicknesses for these two layers for manipulating the field screening, film quality and interface improvement by annealing study, using several other nanostructures (*e.g.*, Ag-NP, Cu-nanostructure) with optical response, and variation of their sizes including that of Au-NPs. Henceforth, the present study predicts that such architectures with engineered materials and geometries may offer superior memory effects with highly improved endurance and retention as well as cycle-to-cycle variability and reliability with enormous future prospect as memristive device.

**Acknowledgement**

The authors would like to acknowledge WBDITE, DST PURSE, Center of Excellence (COE), and Centre for Research in Nanoscience and Nanotechnology (CRNN), University of Calcutta, for providing infrastructural support to conduct this work. Further, the authors acknowledge Mr. Subrata Mandal and Mr. Nilayan Paul of the Department of Electronic Science, University of Calcutta, for the FESEM images, and Mr. Nabakumar Rana of the Department of Physics, University of Calcutta, for the XRD experiments. The authors are also grateful to Dr. Subhrajit Sikdar of the Department of Electronic and Electrical Engineering, The University of Sheffield, for fruitful discussions.